\documentclass[12pt]{article}
\usepackage{times,mathptmx}
\usepackage{graphicx}
\usepackage{bm}

\newcommand{\Elf}{{\cal E}}
\newcommand{\ELF}{\vec{\cal E}}
\newcommand{\mgf}{{\cal B}}
\newcommand{\MGF}{\vec{\cal B}}
\newcommand{\AVP}{\vec{\cal A}}

\newcommand{\rmi}{{\rm i}}
\newcommand{\rme}{{\rm e}}
\newcommand{\rmd}{{\rm d}}

\begin{document}
\title{Electron drift orbits in crossed electromagnetic fields and the quantum Hall effect.}

\author{Tobias Kramer\\
Department of Physics\\
Harvard University\\
17 Oxford Street\\
Cambridge, MA 02138, USA.\\
Email: tobias.kramer@mytum.de}
\date{Updated version, December 4, 2005}

\maketitle

\begin{abstract}
The classical drift motion of electrons in crossed electric and magnetic fields provides an interesting example of a system with an on average constant velocity --  despite the presence of an electric field. This drift-velocity depends solely on the ratio of the electric and magnetic fields and not on the initial momentum of the electron. The present work describes the quantum-mechanical version of this drift-motion, which differs drastically from the classical result: The drift becomes dependent on the energy and a quantization of the transport occurs. The results bear implications for the theory of the quantum Hall effect: Current theories neglect  the electric Hall-field (which is perpendicular to a magnetic field) and thus do not include the quantization due to the crossed-field geometry. I will discuss why it is not possible to eliminate the electric field and how one can explain the quantization in crossed fields in a semiclassical picture. These results make it possible to construct an alternative theory of the quantum Hall effect (see also the detailed description in \cite{Kramer2005c}).
\end{abstract}

\section{Electrons in homogeneous electric and magnetic fields.}

A purely electric field leads to a uniform acceleration of a charged particle, whereas a purely magnetic field forces the particle on a circular path. The Hamiltonian for time-independent electric and magnetic fields can be written as
\begin{equation}\label{eq:HamiltonExB}
H={\left[\bm{p}-\frac{e}{c}\AVP(\bm{r})\right]}^2/(2m)
  -e\ELF\cdot\bm{r},
\end{equation}
where
\begin{equation}
\AVP=\frac{1}{2}\MGF\times\mathbf{r},\qquad
\MGF=(0,0,\mgf),\qquad
\ELF=\ELF_\parallel+\ELF_\perp,\qquad
{\rm and}\qquad
\ELF_\perp\cdot\MGF=0.
\end{equation}
The following discussion assumes $\ELF_\parallel=0$, which is equivalent to a orthogonal-field configuration. Then, it is possible to separate the Hamiltonian into two parts
\begin{equation}
H=p_z^2/(2m)+H_\perp,
\quad
H_\perp=
{\left[\bm{p}-\frac{e}{c}\AVP(\bm{r})\right]}^2/(2m)
-e\ELF_\perp y,
\end{equation}
where the electric field is aligned along the $y$-axis. The part $H_z$ will not be considered, since I am only interested in the electron motion along the plane perpendicular to the magnetic field. This geometry is realized experimentally in two-dimensional electron gases. The classical equations of motion are readily derived:
\begin{equation}
\bm{r}(t)=
\frac{1}{\mgf^2 e}
\left(
\begin{array}{cc}
-\mgf p_y(0)         & \mgf p_x(0)-\Elf_y m\\
\mgf p_x(0)-\Elf_y m & \mgf p_y(0)
\end{array}
\right)
\left[\begin{array}{c}
\cos(e\mgf t/m)\\
\sin(e\mgf t/m)
\end{array}\right]
+
\left[
\begin{array}{c}
\frac{\Elf_y}{\mgf}t+\frac{p_y(0)}{\mgf e}\\
\frac{m\Elf_y}{\mgf^2 e}-\frac{p_x(0)}{\mgf e}
\end{array}
\right].
\end{equation}
The initial position and velocity vectors are given by $\bm{r}(0)=\bm{o}$ and $\dot{\bm{r}}(0)=\bm{p}(0)/m$. The most remarkable property of the motion is the on average (over one period $T=2\pi m/(eB)$) constant drift-velocity:
\begin{equation}\label{eq:DriftVelocity}
\bm{v_d}=\frac{1}{T}\int_{t}^{t+T}{\rm d}t'\,
\dot{\bm{r}}(t')=(\ELF\times\MGF)/\mgf^2.
\end{equation}
The drift-velocity $\bm{v_d}$ is also independent of the initial velocity $\dot{\bm{r}}(0)$. Every electron that starts at $\bm{r}(0)=\bm{o}$ will propagate.

\subsection{The classical Hall effect.}

The constant drift-velocity has important consequences for the transport of electrons in a solid which is placed in a magnetic field. In a classical Hall experiment deflected electrons form an electric field along the edges of a metal. The conducting electrons propagate in the presence of this electric Hall field, which can be used to determine the carrier-density in the sample \cite{Hall1879a}.
Completely neglecting scattering events, we can extract the basic relation between the classical current $\bm{J}$
\begin{equation}
\bm{J}=N e \bm{v_d},
\end{equation}
($N$ denotes the electron density, $e$ the electronic charge) and the resistivity tensor $\bm{\rho}$ (or it's inverse, the conductivity tensor $\bm{\sigma}$) from Ohm's law:
\begin{equation}\label{eq:ConductivityTensor}
\bm{J}=\bm{\rho}^{-1}\cdot\ELF
\quad\Rightarrow\quad
\bm{\rho}^{-1}=
\bm{\sigma}=
\frac{N e}{\mgf}
\left(
\begin{array}{cc}
0 & -1\\
1 & 0
\end{array}
\right).
\end{equation}
One obtains a finite value of the conductivity even in the absence of any scatterers. Without crossed fields, scattering events are used to establish an effective friction force and thus a finite classical conductivity.

\subsection{The quantum Hall effect.}

Current theories of the quantum version of the Hall effect are not based on the Hamiltonian (\ref{eq:HamiltonExB}), but rather on
\begin{equation}\label{eq:HamiltonDisorder}
H_{\rm lattice,disorder}=
{\left[\bm{p}-\frac{e}{c}\AVP(\bm{r})\right]}^2/(2m)
+V_{\rm LD}(\bm{r}),
\end{equation}
where $V_{\rm LD}(\bm{r})$ denotes a periodic lattice potential and possibly uncorrelated disorder potentials (which are often assumed to disappear on the average: $\int\rmd\bm{r}\, V_{\rm LD}(\bm{r})=0$). 
This Hamiltonian differs from the classical Hall Hamiltonian by the omission of the electric Hall field. The disorder potential becomes an essential part of the description and the appearance of a quantized conductivity is linked to the presence of a fluctuating potential-landscape $V_{\rm LD}$ \cite{Hajdu1994a,KramerB2003a}.

In this contribution I do not follow this road. Instead I consider the implications of a quantization of the original Hall-Hamiltonian (\ref{eq:HamiltonExB}). I show that under appropriate boundary conditions this Hamiltonian produces a quantized conductivity --- even without disorder potentials. The inclusion of imperfections is of course desirable, but beyond the scope of this contribution.

\section{Is the classical drift-transport identical to the quantum-mechanical propagation?}

This question is of crucial importance for the theoretical description of a quantum Hall system. If the answer would be yes, one could use the same derivation as for the classical Hall effect and would obtain an unchanged conductivity tensor: No quantization occurs, and the theory of a quantized transport must be based on different effects. Two arguments are used to claim that the quantum and classical result for the drift-motion are equal:
\begin{enumerate}
\item
Many textbooks and reviews describe a translationally invariant system \cite{Imry2002a,Prange1987a,Murthy2003a}. It is correctly noticed that in a frame moving with the classical drift velocity, the electric field vanishes. The presumed implications are that the electric field cannot influence the current, since it is possible to eliminate it by a going to a moving coordinate-system. However, translational invariance must be broken, since the quantized current is observed. The only way to achieve the breaking of invariance in this approach is due to the presence of disorder.\\
~\\
\textbf{But:} 
Translational invariance is never realized in a Hall system. Contacts are present in every direction and electrons will enter the system at the fixed gate-positions. Secondly, it is not enough to transform the already emitted electron. Instead one has to transform the electron-emitting contacts, too. It is well know in atomic physics that the emittance rate of a moving source in a magnetic field changes drastically (new experimental data and a theoretical description are presented in \cite{Yukich2003a}). It is plainly impossible to map the crossed-field problem to a two-dimensional electron-gas in a purely magnetic field without accounting for that motion and the presence of an external current.
\item
Another argument uses the quantum-mechanical expectation value of the momentum operator between eigenstates $|\psi\rangle$ of the Hamiltonian (\ref{eq:HamiltonExB}) \cite{Yoshioka2002a}:
\begin{equation}\label{eq:DriftExpectation}
\left\langle\psi\left|\frac{\bm{p}-\frac{e}{c}\AVP(\bm{r})}{m}\right|\psi\right\rangle=\bm{v_d}.
\end{equation}
This expectation value is indeed identical to the classical time-averaged drift velocity.\\
~\\
\textbf{But:}
In a quantum system with contacts, the expectation value of the momentum alone does not determine the current. A current is only possible if there is a free state available, which can be occupied by the electron. This is the basic idea of every band-model in solid-state physics. One calculates the density of states $n(E)$ of the system and then one fills up the available states according to Pauli's exclusion principle. For a non-interacting Fermi-gas at zero temperature, the upper limit is determined by the Fermi energy. The current is the product of the expectation value with the emission rate (for details, see \cite{Kramer2003d}, Sec.~5.5).
\end{enumerate}
\section{The density of states in crossed fields.}

Surprisingly, the purely quantum-mechanically density of states can be linked to the classical equations of motion \cite{Berry1972a}. Without any external fields, the density of states in two-dimensions becomes a constant, whereas a purely magnetic field leads to a series of discrete Landau levels.

For a non-zero electric field, this is no longer true and the local density density of states (LDOS) becomes a smooth function of the energy \cite{Kramer2003a}. Interestingly, between two former Landau levels the LDOS is extremely suppressed and thus a transport is not possible. This is in sharp contrast to the classical result, in which electrons propagate independently of their initial energy. Even more surprising, another suppression of the LDOS happens inside a former Landau level, which leads to the possibility of fractionally filled levels. These regions of strong transport blocking are transformed into plateaus of constant conductivity \cite{Kramer2003b}.

Notice that the construction of a global current pattern from the LDOS is subject to a coarse-graining procedure, which involves a model of the spatial variation of the Fermi-level, for details see \cite{Kramer2003d}. Here, I just consider a point-like contact at the origin $\bm{o}$, that might already serve as a model for an array of point-contacts with adjusted Fermi levels \cite{Kramer2003a,Kramer2003d}.

The basic idea for a semiclassical interpretation of the LDOS in crossed fields is given by closed orbit theory \cite{Berry1972a,Peters1993a}, which I will apply now.

\subsection{Closed orbits in the classical picture.}

\begin{figure}
\includegraphics[width=1.0\textwidth]{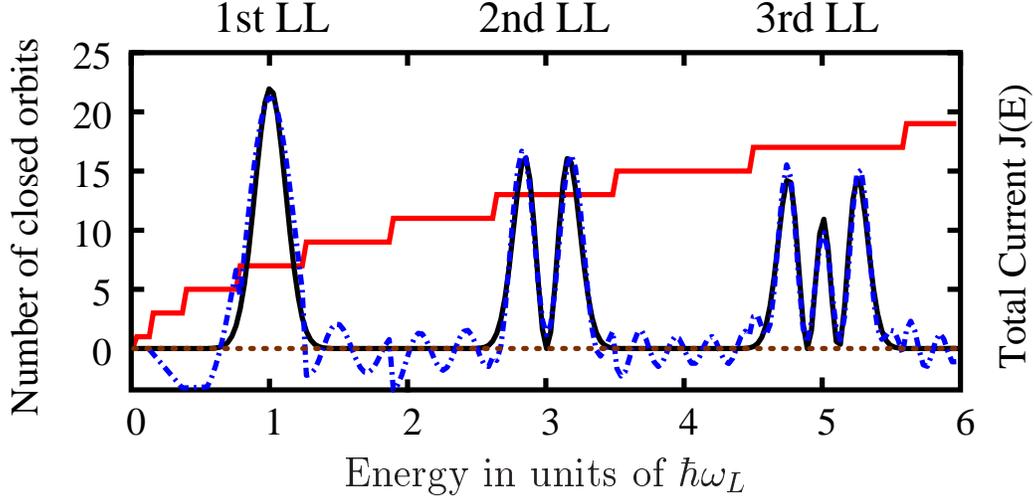}
\caption{
Density of states in crossed electric and magnetic fields. The curves show the quantum (solid) and semiclassical result (dashed line). The staircase structure denotes the count of closed orbits. Crossed electric field $\Elf=4000$~V/m and magnetic field $\mgf=5$~T.
\label{fig:compare}}
\end{figure}

Classical orbits, which lead from the origin back to the origin, are best found by using the classical action. Each classical trajectory must be a extremal point of the reduced classical action $S_{\rm cl}$:
\begin{equation}\label{eq:SaddlePoints}
\frac{\partial S_{\rm cl}(\bm{o},t|\bm{o},0)}{\partial t}+E=0.
\end{equation}
For crossed fields, the classical action is given by
\begin{equation}\label{eq:SclExB}
S_{\rm cl}(\bm{o}, t|\bm{o},0)=
- \frac{m}{2} v_D^2 t
+ \frac{m\omega_L}2 \cot(\omega_L t) v_D^2 t^2,
\end{equation}
where $\omega_L=e\mgf/(2m)$ denotes the Larmor frequency. In Fig.~\ref{fig:compare} I show the number of stationary points as a function of the energy $E$. 

\subsection{The density of states from the propagator.}

The classical action is an important ingredient of the quantum-mechanical time-evolution operator. It is possible to relate the local density of states with the propagator via \cite{Berry1972a,Gutzwiller1990a,Kramer2003a}
\begin{equation}
\label{eq:DOSIntegralExB}
n_{\ELF\times\MGF}^{(2D)}(\mathbf o;E) = \frac{1}{2\pi\;\hbar} \int_{-\infty}^\infty \rmd T\, \rme^{\rmi ET/\hbar}\, K_\perp(\mathbf o, T|\mathbf o,0),
\end{equation}
where the time-dependent propagator is given by
\begin{equation}
K_\perp(\bm{o}, t|\bm{o},0) = \frac{m\omega_L}{2\pi\rmi\;\hbar\sin(\omega_L t)}
\exp \left\{ \frac\rmi\hbar S_{{\rm cl}}(\bm{o}, t|\bm{o},0)\right\}.
\end{equation}
For the case of crossed fields, this expression can be evaluated analytically in terms of harmonic oscillator eigenstates in energy space \cite{Kramer2003a}.
 
\subsection{Comparison of semiclassical and quantum result.}
\begin{figure}
\includegraphics[width=0.7\textwidth]{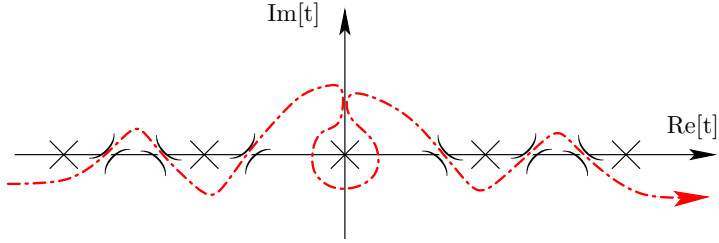}
\caption{
Principal structure of the classical action in the complex time-plane for $N=3$ saddle points. The dashed line denotes the integration path. Singularities are denoted by $\times$ and saddle points by $)($. Note that the singularity at the origin arises from the prefactor in the propagator and not from the classical action at $t=0$.
\label{fig:integrationpath}}
\end{figure}
An asymptotic evaluation of the integral~(\ref{eq:DOSIntegralExB}) provides the link between closed orbits and the density of states. The original path of integration follows the real time-axis. An analytic continuation of the propagator makes it possible to deform this path of integration to the one sketched in Fig.~\ref{fig:integrationpath}. This path passes through saddle points of the exponent (denoted by $)($ in the figure) using the paths of steepest descent. The singularities in the integrand at times $t=n/(\omega_L \pi)$ (denoted by $\times$) are avoided. The only contribution of a singularity comes from $t=0$, which may be evaluated by the residue-theorem:
\begin{equation}
I_{{\rm origin}}
=\frac{1}{2\pi\;\hbar}\oint\rmd t\, 
\rme^{\rmi E t/\hbar}\, K_\perp(\mathbf o, t|\mathbf o,0)
=\frac{m}{2\pi\;\hbar^2}.
\end{equation}
Combining this value with the contributions from the $N$ saddle-points yields the semiclassical result:
\begin{equation}
n_{sc,\ELF\times\MGF}^{(2D)}=I_{{\rm origin}}+2\,{\rm Re}\left[\frac{m\omega_L}{4\pi^2\rmi\;\hbar^2}\sum_{k=1}^N
\frac{\rme^{\rmi E t_k/\hbar+\rmi S_{cl}(\bm{o},t_k|\bm{o},0)/\hbar+\rmi\pi{\rm  sgn}[\ddot{S}_{cl}(\bm{o},t_k|\bm{o},0)]/4}}{\sin(\omega_L t_k)\,\sqrt{|\ddot{S}_{cl}(\bm{o},t_k|\bm{o},0)|/(2\pi\;\hbar)}}
\right]
\end{equation}
Fig.~\ref{fig:compare} compares the semiclassical and quantum result. The agreement between the quantum-mechanical oscillator functions and the interfering classical orbits is striking: properly weighted classical trajectories lead to a quantization into separated levels with a substructure. Note that the number of trajectories is not changed in a single level. Only the relative phase modulates the LDOS.

\section{Conclusion.}

I have shown, that the transport-blocking in crossed fields can be related to the interference of classical trajectories. The blocking can be viewed as a destructive interference of multiple closed trajectories. This non-classical effect leads to a quantization of the quantum-mechanical current emitted from a point contact and makes it possible to construct a quantum theory of the Hall effect based on  Eq.~(\ref{eq:HamiltonExB}), see i.e.\ \cite{Kramer2005c}.

\section{Acknowledgements.}

This work benefited greatly from discussions with M.~Kleber, P.~Kramer, and C.~Bracher. The invitation by M.~Moshinsky for a stay in Mexico with financial support from CONACyT, and the invitation by G.S.~Pogosyan, L.E.~Vicent, and K.B.~Wolf to present this work at the 25th ICGTMP are gratefully acknowledged.

\providecommand{\url}[1]{#1}

\end{document}